\documentstyle[12pt,epsf]{article}
\newcommand{\lsim}{\mathrel{\raisebox{-.6ex}{$\stackrel{\textstyle<}{\sim}$}}}
\newcommand{\gsim}{\mathrel{\raisebox{-.6ex}{$\stackrel{\textstyle>}{\sim}$}}}

\def\beq{\begin{equation}}
\def\eeq{\end{equation}}
\def\bea{\begin{eqnarray}}
\def\eea{\end{eqnarray}}

\begin{document}

\thispagestyle{empty}

\font\fortssbx=cmssbx10 scaled \magstep2
\hbox to \hsize{
\hbox{\fortssbx University of Wisconsin - Madison}
      \hfill$\vcenter{
\hbox{\bf MADPH-98-1070}
\hbox{\bf VAND-TH-98-11}
\hbox{\bf AMES-HET-98-10}
       \hbox{July 1998}}$ }

\vspace{.5in}

\begin{center}
{\bf GENERALIZED NEUTRINO MIXING\\
FROM THE ATMOSPHERIC ANOMALY}\\
\vskip 0.7cm
{V. Barger$^1$, T.J. Weiler$^2$, and K. Whisnant$^3$}
\\[.1cm]
$^1${\it Department of Physics, University of Wisconsin, Madison, WI
53706, USA}\\
$^2${\it Department of Physics and Astronomy, Vanderbilt University,
Nashville, TN 37235, USA}\\
$^3${\it Department of Physics and Astronomy, Iowa State University,
Ames, IA 50011, USA}\\
\end{center}

\smallskip

\begin{abstract}

We determine the neutrino mixing and mass parameters that are allowed by
the Super--Kamiokande atmospheric neutrino data in a three--neutrino
model with one mass--squared difference contributing to the
oscillations. We find that although $\nu_\mu\rightarrow\nu_\tau$
oscillations are favored, $\nu_\mu\rightarrow\nu_e$ oscillations with
amplitude as large as 0.18 are allowed even after accounting
for the limit from the CHOOZ reactor experiment. The range of allowed
parameters permit observable $\nu_\mu\leftrightarrow\nu_e$ and
$\nu_e\rightarrow\nu_\tau$ oscillations in future long--baseline
experiments.

\end{abstract}

\thispagestyle{empty}
\newpage

{\it\underline{Introduction}.} It was suggested long ago~\cite{oldatmos}
that the atmospheric neutrino anomaly~\cite{atmos} could be explained by
the oscillation of muon neutrinos and antineutrinos into another
neutrino species. This interpretation has been confirmed by the zenith
angle dependence measured by the Super--Kamiokande (SuperK)
experiment~\cite{SuperK}. Neutrino oscillations can also be invoked to
separately explain the solar neutrino deficit~\cite{solar,newsolar} and
the results of the LSND experiment~\cite{LSND}. Because confirmation of
the LSND results awaits future experiments and recent measurements in
the KARMEN detector exclude part of the LSND allowed
region~\cite{LSNDconfirm}, a conservative approach is to assume that
oscillations need only account for the solar and atmospheric data. Then
the two mass--squared difference scales in a three--neutrino model are
sufficient to describe the data. Interest in the implications for models
of the atmospheric neutrino anomaly has recently
intensified~\cite{newatmos,bpww,maxosc}.  An attractive possibility is
that both the atmospheric $\nu_\mu$ and solar $\nu_e$ oscillate
maximally or near--maximally at the $\delta m^2_{atm}$ and $\delta
m^2_{sun}$ scales, respectively~\cite{bpww,maxosc}.

In this letter we use the recent Super--Kamiokande atmospheric
neutrino data~\cite{SuperK} to determine the allowed values for the
general three--neutrino mixing matrix under the assumption that one
mass--squared difference, $\delta m^2_{atm}$, explains the atmospheric
neutrino oscillations. In our scenario the other mass--squared
difference, $\delta m^2_{sun}\ll\delta m^2_{atm}$, can explain the solar
neutrino oscillations via either an MSW~\cite{MSW} or vacuum
long--wavelength scenario~\cite{vlw}. We find that although pure
$\nu_\mu\rightarrow\nu_\tau$ oscillations of atmospheric neutrinos are
favored, there exist three--neutrino solutions with non--negligible
$\nu_\mu\leftrightarrow\nu_e$ oscillations, even after
applying the constraints from the CHOOZ reactor
experiment~\cite{CHOOZ}. Consequently
$\nu_\mu\leftrightarrow\nu_e$ and $\nu_e\rightarrow\nu_\tau$
oscillations may be observable in future long--baseline
experiments.

{\it\underline{Oscillation probabilities}.} We begin our analysis with
the survival probability for a given neutrino flavor in a vacuum
\cite{VBreal}
\beq
P(\nu_\alpha\to \nu_\alpha) = 1
- 4 \sum_{k<j} P_{\alpha j} P_{\alpha k} \sin^2 \Delta_{jk} \,,
\label{oscprob}
\eeq
where
\beq
P_{\alpha j} \equiv |U_{\alpha j}|^2 \,,
\label{p}
\eeq
$U$ is the neutrino mixing matrix (in the basis where the
charged--lepton mass matrix is diagonal), $\Delta_{jk} \equiv \delta
m_{jk}^2 \,L/4E = 1.27 (\delta m^2_{jk}/{\rm eV}^2) (L/{\rm km})/(E/{\rm
GeV})$, $\delta m^2_{jk}\equiv m^2_j-m^2_k$, and the sum is over all $j$
and $k$, subject to $k<j$. The matrix elements $U_{\alpha j}$ are the
mixings between the flavor ($\alpha=e,\mu,\tau$) and the mass
($j=1,2,3$) eigenstates, and we assume without loss of generality that
$m_1 < m_2 < m_3$. The solar oscillations are driven by
$|\Delta_{21}| \equiv \Delta_{sun}$ and the atmospheric oscillations are
driven by $|\Delta_{31}| \simeq |\Delta_{32}| \equiv \Delta_{atm} \gg
\Delta_{sun}$.

The off-diagonal vacuum oscillation probabilities of this three-neutrino
model are
\begin{eqnarray}
P(\nu_e\rightarrow\nu_\mu) &=&
4\,P_{e3} P_{\mu 3} \sin^2\Delta_{atm}
-4 Re\{ U_{e1} U_{e2}^* U_{\mu 1}^* U_{\mu 2}\} \sin^2\Delta_{sun}
-2\,J\sin 2\Delta_{sun}\,,
\label{pemu} \\
P(\nu_e\rightarrow\nu_\tau) &=&
4\,P_{e3} P_{\tau 3} \sin^2\Delta_{atm}
-4 Re\{U_{e1} U_{e2}^* U_{\tau 1}^* U_{\tau 2}\} \sin^2\Delta_{sun}
+2\,J\sin 2\Delta_{sun}\,,
\label{petau} \\
P(\nu_\mu\rightarrow\nu_\tau) &=&
4\,P_{\mu 3} P_{\tau 3} \sin^2\Delta_{atm}
-4 Re\{ U_{\mu 1} U_{\mu 2}^* U_{\tau 1}^* U_{\tau 2}\} \sin^2\Delta_{sun}
-2\,J\sin 2\Delta_{sun}\,,
\label{pmutau}
\end{eqnarray}
where the $CP$--violating ``Jarlskog invariant'' \cite{jarlskog}
is $J = \sum_{k,\gamma} \epsilon_{ijk} \epsilon_{\alpha\beta\gamma}
Im\{ U_{\alpha i} U_{\alpha j}^* U_{\beta i}^* U_{\beta j}\}$ for any
$\alpha$, $\beta$, $i$, and $j$ (e.g., $J = Im\{ U_{e2} U_{e3}^*
U_{\mu 2}^* U_{\mu 3}\}$ for $\alpha=e$, $\beta=\mu$, $i=2$, and $j=3$).
The $CP$--odd term changes sign under reversal of the oscillating flavors.
We note that the $CP$--violating probability at the atmospheric scale is
suppressed to order $\delta m^2_{sun}/\delta m^2_{atm}$, the leading
term having cancelled in the sum over the two light--mass states; thus,
$P(\nu_\alpha \rightarrow \nu_\beta) = P(\nu_\beta \rightarrow
\nu_\alpha)$ at the atmospheric scale.

{\it\underline{Fit to atmospheric neutrino data}.} For the $L/E$ values of
the atmospheric and long--baseline experiments, $\Delta_{sun}$ can be
neglected, and the vacuum oscillation probabilities become simply
\begin{eqnarray}
P(\nu_\mu\rightarrow\nu_\mu) &=&
1 - 4\,P_{\mu 3} (1 - P_{\mu 3}) \sin^2\Delta_{atm} \,,
\label{pmumu2} \\
P(\nu_e\rightarrow\nu_e) &=& 1 -
4\,P_{e3} (1 - P_{e 3}) \sin^2\Delta_{atm} \,,
\label{pee2} \\
P(\nu_e\leftrightarrow\nu_\mu) &=&
4\,P_{e3} P_{\mu 3} \sin^2\Delta_{atm} \,,
\label{pemu2} \\
P(\nu_e\leftrightarrow\nu_\tau) &=&
4\,P_{e3} (1 - P_{e3} - P_{\mu 3}) \sin^2\Delta_{atm} \,,
\label{petau2} \\
P(\nu_\mu\leftrightarrow\nu_\tau) &=&
4\,P_{\mu 3} (1 - P_{e3} - P_{\mu 3}) \sin^2\Delta_{atm} \,.
\label{pmutau2}
\end{eqnarray}
In Eqs.~(\ref{petau2}) and (\ref{pmutau2}) we have used
the unitarity condition $P_{\tau 3} = 1 - P_{e3}
- P_{\mu 3}$. Thus for oscillations at the atmospheric scale
there are only two independent mixing matrix parameters,
e.g., $P_{e3}$ and $P_{\mu 3}$, that are relevant. All predictions for
atmospheric and long--baseline experiments are completely determined by
the three parameters $\delta m^2_{atm}$, $P_{e3}$, and $P_{\mu 3}$.
We define the oscillation amplitudes $A^{\mu\not\mu}_{atm}$,
$A^{e \not e}_{atm}$, $A^{\mu e}_{atm}$, $A^{e\tau}_{atm}$, and
$A^{\mu\tau}_{atm}$, as the coefficients of the $\sin^2 \Delta_{atm}$
terms in Eqs.~(\ref{pmumu2})--(\ref{pmutau2}), respectively.
The parameters $P_{e3}$ and $P_{\mu3}$ can then be determined from the
atmospheric neutrino data by the relations
\beq
N_\mu/N^o_\mu = \alpha \left[(1 - \left< S \right> \, A^{\mu\not\mu}_{atm})
+ r \, \left< S \right> \, A^{\mu e}_{atm} \right] \,,
\label{Rmu}
\eeq
and
\beq
N_e/N^o_e = \alpha \left[ (1 - \left< S \right> \, A^{e\not e}_{atm})
+ r^{-1} \, \left< S \right> \, A^{\mu e}_{atm} \right] \,,
\label{Re}
\eeq
where $N_e^o$ and $N_\mu^o$ are the expected numbers of atmospheric $e$
and $\mu$ events, respectively, $r\equiv N^o_e/N^o_\mu$, $\left< S
\right>$ is $\sin^2\Delta_{atm}$ appropriately averaged,
and $\alpha$ is the overall neutrino flux normalization, which we allow
to vary following the SuperK analysis~\cite{SuperK}. SuperK reports
$N_{\mu}/N_{\mu}^o$ and $N_e/N_e^o$ from a 535 day exposure for eight
different $L/E$ bins \cite{SuperK}. The data were obtained by inferring
an $L/E$ value for each event from the zenith angle $\theta_\ell$ and
energy of the observed charged lepton $E_\ell$ and comparing it to
expectations from a monte carlo simulation based on the atmospheric
neutrino spectrum~\cite{flux} folded with the differential cross
section.

Due to the fact that the charged lepton energy and direction in general
differ from the corresponding values for the incident neutrino (or
antineutrino), the $L/E$ distribution involves substantial smearing. We
estimate this smearing by a monte carlo integration over the neutrino
angle and energy spectrum~\cite{stanevflux} weighted by the differential
cross section. We generate events with
$E_\nu$ and $\theta_\nu$, and determine the corresponding $E_\ell$ and
$\theta_\ell$ for the charged lepton. We bin the events in $L/E_\nu$,
using $\theta_\ell$ to determine $L$ and an estimated neutrino energy
inferred from the average ratio of lepton momentum to neutrino energy,
$E_\nu^{\rm est} = E_\ell \left< E_\nu/E_\ell \right>$, analogous to the
SuperK analysis~\cite{SuperK}. We then calculate a value for $\left<
\sin^2\Delta_{atm} \right>$ for each $L/E$ bin for a given value of
$\delta m^2_{atm}$. Finally we make a fit to Eqs.~(\ref{Rmu}) and
(\ref{Re}) to determine $P_{e3}$, $P_{\mu3}$, $\delta m^2_{atm}$, and
$\alpha$~\cite{bilenky}. Without loss of generality we take $\delta
m^2_{atm}$ to be positive.

Our best fit values for the four parameters are
\begin{eqnarray}
\delta m^2_{atm} &=& 2.8\times10^{-3}{\rm~eV}^2 \,,
\label{dm2best}\\
P_{e3} &=& 0.00 \,,
\label{Ue3best}\\
P_{\mu 3} &=& 0.50 \,,
\label{Umu3best}\\
\alpha &=& 1.16 \,,
\label{alphabest}
\end{eqnarray}
with $\chi^2_{min}=7.1$ for 12 degrees of freedom. This best fit is
close to the result of the SuperK simulation that assumed
only two-neutrino oscillations. In Fig.~1a we show the
allowed region for $P_{\mu 3}$ versus $P_{e3}$ for $\alpha=1.16$ and
$\delta m^2_{atm} = 2.8\times10^{-3}$~eV$^2$. Although $P_{e3}=0$ is
favored, small nonzero values are allowed, which permit some $\nu_\mu
\leftrightarrow \nu_e$ and $\nu_e \rightarrow \nu_\tau$ oscillations
of atmospheric neutrinos. In Fig.~1b we show the allowed region for the
overall flux normalization $\alpha$ versus $\delta m^2_{atm}$ for
$P_{e3}=0$ and $P_{\mu 3}=0.50$.

Another limit on $P_{e3}$ comes from the CHOOZ reactor experiment
\cite{CHOOZ} that measures $\bar\nu_e$ disappearance
\beq
A^{e\not e}_{atm} = 4 P_{e3} (1 - P_{e3}) \lsim 0.2 \,,
\label{choozlimit}
\eeq
which applies for $\delta m^2_{atm} \gsim 2\times10^{-3}$~eV$^2$. The
exact limit on $A^{e\not e}_{atm}$ varies with $\delta m^2_{atm}$, and
for $\delta m^2_{atm} < 10^{-3}$~eV$^2$ there is no limit at all. For
$\delta m^2_{atm} = 2.8\times10^{-3}$~eV$^2$ and $\alpha=1.16$, $P_{e3}$
is constrained to be less than 0.04. The result of this additional
constraint is shown in Fig.~1a. In Fig.~1c we show the effect of the
CHOOZ constraint on the allowed region of $P_{e3}$ versus $\delta
m^2_{atm}$ for $\alpha=1.16$ and $P_{\mu 3}=0.50$.


Varying over the entire parameter space, and imposing the CHOOZ
constraint, the ranges of allowed values at 68\% (95\%) C.L. are
\begin{eqnarray}
0.8 \, (0.5) &\leq& \delta m^2_{atm}/(10^{-3}{\rm~eV}^2) \leq 7.9 \, (10.0) \,,
\label{dm2range}\\
0.00 \, (0.00) &\leq& P_{e3} \leq 0.05 \, (0.08) \,,
\label{Ue3range}\\
0.29 \, (0.25) &\leq& P_{\mu 3} \leq 0.71 \, (0.75) \,,
\label{Umu3range}\\
1.07 \, (1.04) &\leq& \alpha \leq 1.24 \, (1.28) \,.
\label{alpharange}
\end{eqnarray}
The allowed ranges of some related oscillation parameters are
\begin{eqnarray}
0.29 \, (0.25) &\leq& P_{\tau 3} \leq 0.71 \, (0.75)
\label{ptau3}\\
0.83 \, (0.75) &\leq& A^{\mu\not\mu}_{atm} \leq 1.00 \, (1.00) \,,
\label{amumurange}\\
0.00 \, (0.00) &\leq& A^{e\not e}_{atm} \leq 0.18 \, (0.29) \,,
\label{aeerange}\\
0.00 \, (0.00) &\leq& A^{\mu e}_{atm} \leq 0.11 \, (0.18) \,,
\label{amuerange}\\
0.00 \, (0.00) &\leq& A^{e\tau}_{atm} \leq 0.10 \, (0.16) \,,
\label{aetaurange}\\
0.82 \, (0.74) &\leq& A^{\mu\tau}_{atm} \leq 1.00 \, (1.00) \,.
\label{amutaurange}
\end{eqnarray}
Although $\nu_\mu\to\mu_\tau$ oscillations are strongly favored,
$\nu_\mu\leftrightarrow\nu_e$ are allowed with amplitude as large as 0.18.

We do not consider matter effects in the above analysis. For the $\delta
m^2_{atm}$ favored by our fit, matter effects are small for the sub-GeV
neutrinos that constitute most of the data. Also, as evidenced by our
fit, the dominant oscillation is $\nu_\mu \rightarrow \nu_\tau$, which
is not affected by matter. However, matter effects could be important
for neutrinos with smaller $\delta m^2_{atm}/E$, i.e., especially for
multi-GeV data in solutions with $\delta m^2_{atm}\lsim 10^{-3}\rm\,eV^2$,
which could be observable as more data becomes available at higher energies.

{\it\underline{New physics predictions}.} The new physics predicted when
$P_{e3} \ne 0$ is $\nu_e\rightarrow\nu_\tau$ oscillations with leading
probability (i.e., the $\Delta_{atm}$ terms) given by
Eq.~(\ref{petau2}). From Eq.~(\ref{aetaurange}) the maximal amplitude
$A^{e\tau}_{atm}$ for these oscillations is 0.16 at 95\%~C.L. Figure~2
shows the allowed values of $A^{e\tau}_{atm}$ versus $A^{\mu e}_{atm}$
for $\delta m^2 = 2.8\times10^{-3}$~eV$^2$ and $\alpha=1.16$; the effect
of the CHOOZ constraint is also shown. The $\nu_e\rightarrow\nu_\tau$
oscillations could be observed by long--baseline neutrino experiments
with proposed high intensity muon sources \cite{geer,bww97,bpww2}, which
can also make precise measurements of $\nu_\mu \leftrightarrow \nu_e$
and $\nu_\mu \rightarrow \nu_\tau$ oscillations. Sensitivity to
$A^{e\tau}_{atm} (\delta m^2_{atm} / {\rm eV}^2)^2 > 2.5\times10^{-9}$
is expected \cite{geer} for the parameter ranges of interest here; for
$\delta m^2 = 2.8\times10^{-3}$~eV$^2$, $A^{e\tau}_{atm}$ could be
measured down to $3\times10^{-4}$. The measurement of
$\nu_e\rightarrow\nu_\tau$ and $\nu_\mu \leftrightarrow \nu_e$
oscillations in such a long--baseline experiment would test the
three-neutrino model.

{\it\underline{Discussion}.}  Further measurements of atmospheric
neutrinos will more precisely determine the parameters $P_{e3}$, $P_{\mu
3}$, and $\delta m^2_{atm}$.  The MINOS~\cite{MINOS}, K2K~\cite{K2K},
and ICARUS~\cite{ICARUS} experiments are expected to test for $\nu_\mu
\rightarrow \nu_e$ and $\nu_\mu \rightarrow \nu_\tau$ oscillations for
$\delta m^2_{atm} > 10^{-3}$~eV$^2$. Together these measurements could
put strong limits on $P_{e3}$, which governs the $\nu_e \rightarrow
\nu_\tau$ oscillations that could be seen in future long--baseline
experiments such as those utilizing a muon storage ring at
Fermilab~\cite{geer}. Full three-neutrino fits including the solar
neutrino data~\cite{osland} can then determine one of the remaining two
independent parameters in the mixing matrix, e.g., $P_{e1}$, using the
$\nu_e$ survival probability for $\Delta_{atm} \gg1$
\beq
P(\nu_e \rightarrow \nu_e) = 1 - 2 P_{e3} (1 - P_{e3})
- 4 P_{e1} (1 - P_{e1} - P_{e3}) \sin^2\Delta_{sun} \,.
\eeq
The considerations in this paper can also be extended to a four--neutrino
model~\cite{bpww2}.

\bigskip
{\it\underline{Acknowledgements}.}
We thank John Learned for stimulating discussions regarding the
Super--Kamiokande atmospheric data, and we thank Sandip Pakvasa for
collaboration on previous related work. We are grateful to Todor Stanev
for providing his atmospheric neutrino flux program.
This work was supported in part by the U.S. Department of Energy,
Division of High Energy Physics, under Grants No.~DE-FG02-94ER40817,
No.~DE-FG05-85ER40226, and No.~DE-FG02-95ER40896, and in part by the
University of Wisconsin Research Committee with funds granted by the
Wisconsin Alumni Research Foundation and the Vanderbilt University
Research Council.

\newpage


%



\begin{figure}
\centering\leavevmode
\epsfxsize=6.5in\epsffile{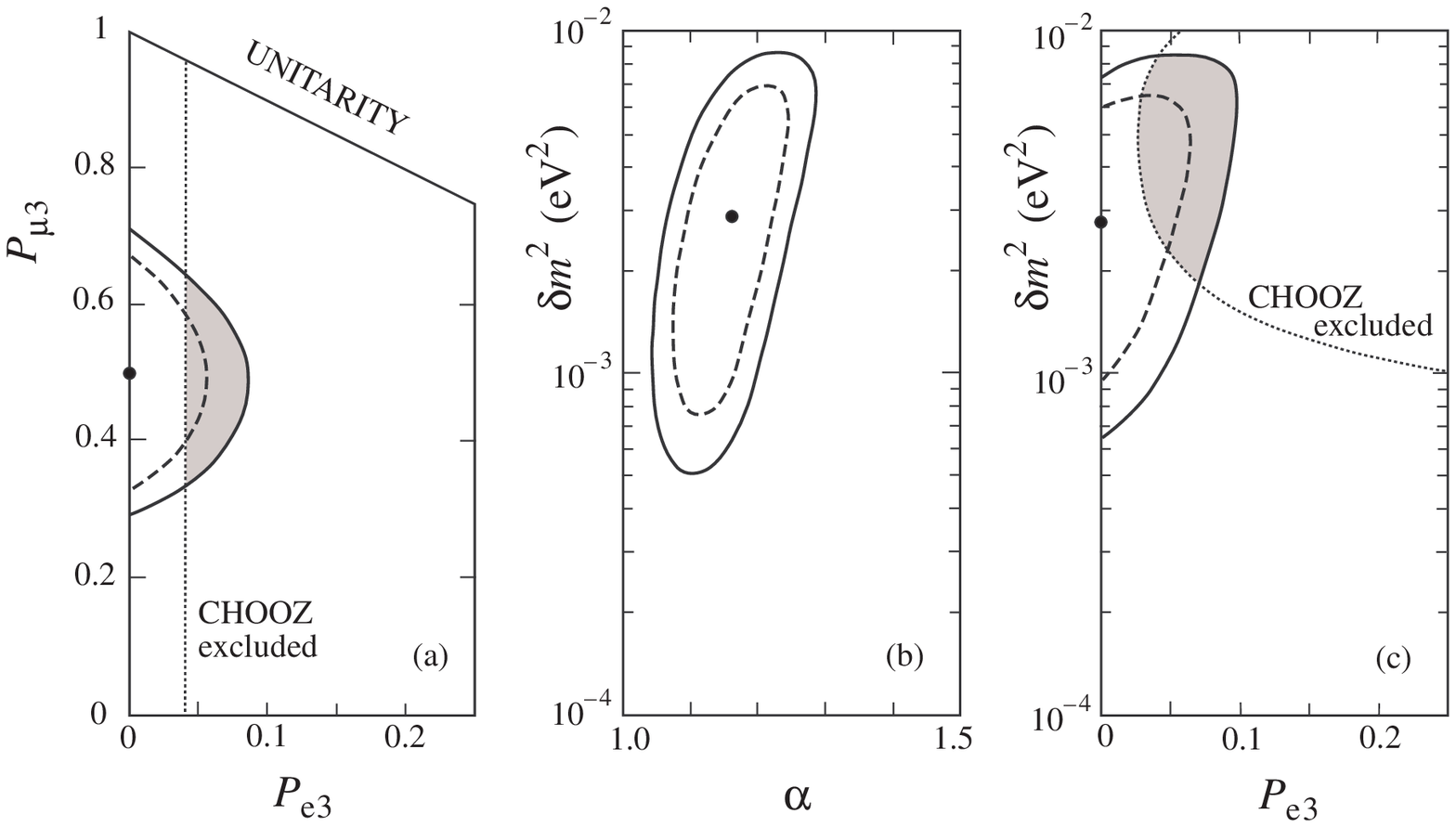}

\caption[]{\label{Fig1} Allowed regions of the four--parameter space
($\delta m^2_{atm}$, $P_{e3}$, $P_{\mu 3}$, $\alpha$) at 68\% (dashed
line) and 95\% C.L. (solid line) for (a) $P_{\mu 3}$ versus $P_{e3}$
with $\delta m^2_{atm}=2.8\times10^{-3}$~eV$^2$ and $\alpha=1.16$, (b)
$\delta m^2_{atm}$ versus $\alpha$ with $P_{e3}=0$ and $P_{\mu 3}=0.50$,
and (c) $\delta m^2_{atm}$ versus $P_{e3}$ with $P_{\mu 3}=0.50$ and
$\alpha=1.16$. The best fit values are indicated by the filled circles.
The CHOOZ constraint~\cite{CHOOZ}, shown by a dotted line, excludes the
shaded region.}
\end{figure}

\clearpage


\begin{figure}
\centering\leavevmode
\epsfxsize=5.5in\epsffile{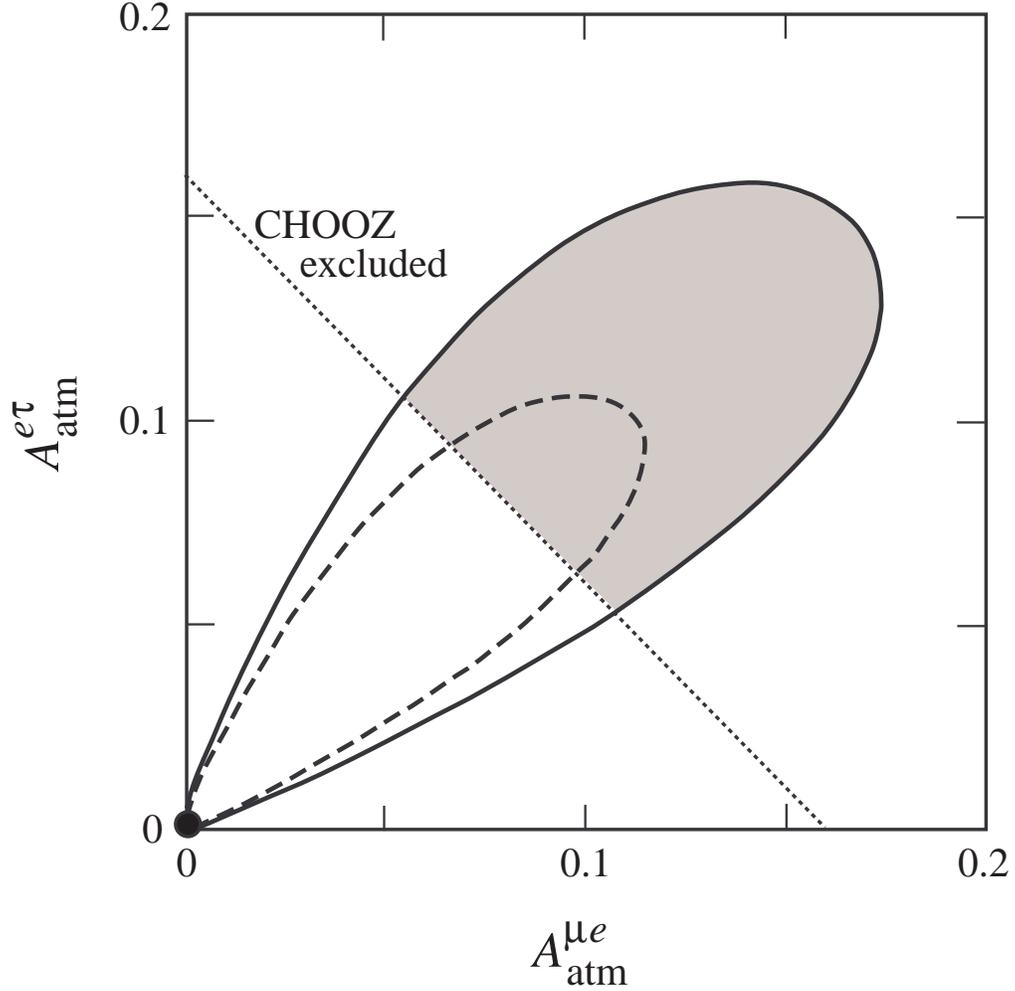}

\caption[]{\label{Fig2} Allowed regions at 68\% (dashed line) and
95\% C.L. (solid line) for the oscillation amplitudes $A_{e\tau}$
versus $A_{\mu e}$ with $\delta m^2=2.8\times10^{-3}$~eV$^2$ and
$\alpha=1.16$. The best fit values are indicated by the filled circle.
The CHOOZ constraint~\cite{CHOOZ}, shown by a dotted line, excludes the
shaded region.}
\end{figure}


\begin{thebibliography}{99}
\bibitem{stanevflux}
The flux used in our calculation was provided by T. Stanev.
\end{thebibliography}
\end{document}